\documentclass[floats,aps,amssymb,nofootinbib,superscriptaddress]{revtex4}
\usepackage{amssymb}
\usepackage{graphics}
\usepackage{amsmath}
\usepackage{amsfonts}
\usepackage{bm}% bold math
\usepackage[]{latexsym}
\usepackage{float}
\usepackage{graphicx}
\usepackage{diagbox}
\usepackage{comment}
\usepackage{xcolor}
\usepackage{xspace}
\usepackage[normalem]{ulem}
\usepackage{dsfont}

\newcommand{\id}{\mathds{1}}
\newcommand\ie{\mbox{\textit{i.\,e.}}\xspace}

\newcommand{\be}{\begin{equation}}\newcommand{\ee}{\end{equation}}
\newcommand{\bea}{\begin{eqnarray}}\newcommand{\eea}{\end{eqnarray}}
\newcommand{\brr}{\begin{array}}\newcommand{\err}{\end{array}}
\newcommand{\bit}{\begin{itemize}}\newcommand{\eit}{\end{itemize}}
\newcommand{\ben}{\begin{enumerate}}\newcommand{\een}{\end{enumerate}}

\newcommand{\ba}{\begin{array}}
\newcommand{\ea}{\end{array}}

\def\lf{\left}

\def\ri{\right}
\def\al{\alpha}

\def\si{\sigma}

\def\1{{_{1}}}\def\2{{_{2}}}

\def\noHe0{:\;\!\!\;\!\!:H_e(0):\;\!\!\;\!\!:}
\def\noHm0{:\;\!\!\;\!\!:H_\mu(0):\;\!\!\;\!\!:}

\def\lf{\left}

\def\ri{\right}

\def\al{\alpha}

\def\si{\sigma}

\def\1{{_{1}}}\def\2{{_{2}}}

\definecolor{nicegreen}{rgb}{0.07, 0.564, 0.04}

\begin{document}
\title{Bell nonlocality in maximal-length quantum mechanics}
%\title{Nonlocality degradation from the cosmological constant}

\author{Pasquale Bosso}
\email[Corresponding author: ]{pbosso@unisa.it}
\affiliation{Dipartimento di Ingegneria Industriale, Universit\`a degli Studi di Salerno, Via Giovanni Paolo II, 132 I-84084 Fisciano (SA), Italy}
\affiliation{INFN, Sezione di Napoli, Gruppo collegato di Salerno, Via Giovanni Paolo II, 132 I-84084 Fisciano (SA), Italy}

\author{Fabrizio Illuminati}
\email{filluminati@unisa.it}
\affiliation{Dipartimento di Ingegneria Industriale, Universit\`a degli Studi di Salerno, Via Giovanni Paolo II, 132 I-84084 Fisciano (SA), Italy}
\affiliation{INFN, Sezione di Napoli, Gruppo collegato di Salerno, Via Giovanni Paolo II, 132 I-84084 Fisciano (SA), Italy}

\author{Luciano Petruzziello}
\email{lupetruzziello@unisa.it}
\affiliation{Dipartimento di Ingegneria Industriale, Universit\`a degli Studi di Salerno, Via Giovanni Paolo II, 132 I-84084 Fisciano (SA), Italy}
\affiliation{INFN, Sezione di Napoli, Gruppo collegato di Salerno, Via Giovanni Paolo II, 132 I-84084 Fisciano (SA), Italy}
\affiliation{Institut f\"ur Theoretische Physik, Albert-Einstein-Allee 11, Universit\"at Ulm, 89069 Ulm, Germany}

\author{Fabian Wagner}
\email{fwagner@unisa.it}
\affiliation{Dipartimento di Ingegneria Industriale, Universit\`a degli Studi di Salerno, Via Giovanni Paolo II, 132 I-84084 Fisciano (SA), Italy}
\affiliation{INFN, Sezione di Napoli, Gruppo collegato di Salerno, Via Giovanni Paolo II, 132 I-84084 Fisciano (SA), Italy}

\date{\today}
\def\be{\begin{equation}}
\def\ee{\end{equation}}
\def\al{\alpha}
\def\bea{\begin{eqnarray}}
\def\eea{\end{eqnarray}}

\begin{abstract}
In this paper, we investigate the consequences of maximal length as well as minimal momentum scales on nonlocal correlations shared by two parties of a bipartite quantum system.
To this aim, we rely on a general phenomenological scheme which is usually associated with the non-negligible spacetime curvature at cosmological scales, namely the extended uncertainty principle.
In so doing, we find that quantum correlations are degraded if the deformed quantum mechanical model mimics a positive cosmological constant.
This opens up the possibility to recover classicality at sufficiently large distances.
\end{abstract}

\maketitle

\section{Introduction}

Both quantum as well as cosmological aspects of gravity have always been notoriously difficult to address rigorously.
Regarding the former, several theoretical inconsistencies as well as a substantial lack of experimental evidence prevent the formulation of a universally accepted theory. Nevertheless, the quest for a cogent and realistic solution to the above conundrums has led to the development of a significant number of valuable models, such as string theory~\cite{Polchinski:1998rq,Polchinski:1998rr}, loop quantum gravity~\cite{Rovelli:2004tv,Rovelli:2014ssa}, causal dynamical triangulations \cite{Ambjorn:2004qm,Loll:2019rdj,Loll:2022ibq} and asymptotically safe quantum gravity~\cite{Reuter:1996cp,Niedermaier:2006wt,Eichhorn:2018yfc}.
%, and doubly special relativity~\cite{Amelino-Camelia:1997ieq,Magueijo:2001cr,Amelino-Camelia:2008aez}.
%On the other hand, great efforts in the fields of cosmology and astroparticle physics are devoted to the search for dark matter and dark energy, themselves affecting a number of unexplained phenomena.
%For instance, it is worth mentioning that dark matter has been proposed to be made up of weakly interacting massive particles~\cite{Bertone:2004pz} such as sterile neutrinos~\cite{Boyarsky:2018tvu}.
On the other hand, cosmological investigations have led to the emergence of unexpected mechanisms and phenomena that cannot find a satisfactory explanation within the currently available models of our Universe. For instance, other than being described by a cosmological constant (whose origin is still argument of intense debate), dark energy continues to be a rather elusive concept, since its theoretical understanding requires the introduction of exotic schemes involving a fifth fundamental interaction (such as quintessence~\cite{Caldwell:1997ii,Carroll:1998zi,Dvali:2001dd}).

To overcome these evident difficulties, %and at the same time get an insight on how Nature truly operates, 
many recent analyses in the context of quantum gravity have employed phenomenological schemes. A similar tendency can also be encountered in the study of cosmological implications on quantum systems.
Specifically, one of 
% those
such schemes consists in suitably deforming the Heisenberg uncertainty relations.
According to the ordinary formula
\be\label{hup}
\Delta x^i\Delta p_j\geq\frac{\hbar}{2}\delta^i_j\,,
\ee
arbitrary distances and momenta can be resolved by resorting to appropriate probes.
The aforesaid scenario, however, no longer holds true 
in presence of
limits to the resolution of length and/or momentum scales, some of which are predicted by quantum gravitational (minimal length) and cosmological (maximal length/minimal momentum) arguments.
In one spatial dimension, the above features are taken into account by modifying Eq.~\eqref{hup} as follows:
\be\label{eur}
\Delta x\Delta p\geq\frac{\hbar}{2}\lf(1+\alpha\Delta x^2+\beta\Delta p^2\ri)\,,
\ee
where the dimensionful $\alpha$ and $\beta$ are referred to as deformation parameters.

The concept of minimal length scale has a long history in quantum gravity \cite{Bronstein36,Mead:1964zz} (for a review see \cite{Garay:1994en}). As of late, it has drawn considerable attention when emerging in super-Planckian string-scattering amplitudes~\cite{Amati:1987wq}. Nevertheless, the existence of a minimal length has found further support in achievements obtained in  loop quantum gravity \cite{Rovelli94} and asymptotic safety \cite{Ferrero:2022hor}, as well as in heuristic arguments and other physical frameworks (for a comprehensive but not exhaustive list of recent developments on this topic, see Refs.~\cite{Maggiore:1993rv,Kempf:1994su,Adler:1999bu,Scardigli:1999jh,Adler:2001vs,Scardigli:2003kr,Das:2008kaa,Jizba:2009qf,Hossenfelder:2012jw,Chen:2014bva,Chen:2014jwq,Scardigli:2016pjs,Todorinov:2018arx,Kumar:2019bnd,Buoninfante:2019fwr,Kanazawa:2019llj,Luciano:2019mrz,Bosso:2020aqm,Buoninfante:2020cqz,Petruzziello:2020een,Buoninfante:2020guu,Petruzziello:2020wkd,Luciano:2021cna,Bosso:2023sxr} and therein).
Motivated by these findings in the quantum regime of the gravitational interaction, the presence of a minimal momentum has been initially proposed as a Born reciprocal of the previous considerations~\cite{Hinrichsen:1995mf,Kempf:1996ss}.
%, but later on 
However, later on it has gained a life on its own as a toy model mimicking the cosmological horizon in non-relativistic quantum mechanics (for further readings, see Refs.~\cite{Bolen:2004sq,Scardigli:2006eb,Park:2007az,Mignemi:2009ji,Zarei:2009nhg,Schurmann:2009zz,CostaFilho:2016wvf,Ong:2018nzk,Schurmann:2018yuz,Dabrowski:2020ixn,Wagner:2021thc,Petruzziello:2021vyf,Illuminati:2021wfq}).
To differentiate between the possible generalizations of the uncertainty relation, in the literature it is customary to address distinct approaches with distinct names. A summary of this nomenclature can be found in Table \ref{table}.

\begin{table}[!h]
    \renewcommand{\arraystretch}{1.2}
    \caption{Different shapes of deformed uncertainty relations}
    \label{table}
    \centering
    \begin{tabular}{|c|c|c|}
        \hline
        \diagbox{Value of $\alpha$}{Value of $\beta$} & $\beta=0$ & $\beta\neq0$ \\
        \hline
        $\alpha=0$ & Heisenberg uncertainty principle (HUP) & Generalized uncertainty principle (GUP) \\
        \hline
        $\alpha\neq0$ & Extended uncertainty principle (EUP) & Extended generalized uncertainty principle (EGUP) \\
        \hline
    \end{tabular}
\end{table}

%\noindent
In the present work, we investigate how quantum nonlocality is influenced by the existence of a minimal momentum or maximal length scale, thus making use of quantum mechanics with EUP. In so doing, we complement the analysis of a recent article~\cite{Bosso:2022ogb}, where a similar investigation has been carried out based on the GUP. Despite this, the crucial difference lies in the fact that the resulting achievements turn out to diverge significantly.

For the problem at hand, a convenient physical scheme to measure the degree of nonlocality is provided by Bell theorem~\cite{Bell:1964kc} in relation to a system of paired spin-1/2 particles. In particular, we resort to the Clauser-Horne-Shimony-Holt (CHSH) version of the Bell-type experiment~\cite{Clauser:1969ny}, which also allows to discuss the implications of the EUP at the level of the maximal violation of Bell's inequality, known as Tsirelson bound~\cite{tsir}.
This threshold quantifies the highest amount of nonlocal correlations that can be reached in the context of standard quantum mechanics, and hence it is reasonable to assume that it is affected by the presence of a non-vanishing minimal momentum.

The manuscript is organized as follows: in Sec. II, we summarize the main features of general EUPs. In Sec. III, we analyze how the employment of the EUP affects the degree of nonlocal correlations for a bipartite system as described by the CHSH version of the Bell theorem. Finally, Sec. IV contains concluding remarks and future perspectives. 

Throughout the whole paper, we use natural units $\hbar=c=1$.

\section{Extended uncertainty principle}

In order to properly introduce the EUP, we shall start from the deformation at the level of the algebra satisfied by the position and momentum operators.
To this aim, we consider the most general, isotropic formulation of the EUP.
The corresponding algebra is given by 
\be\label{eup}
\lf[\hat{x}^i,\hat{x}^j\ri]=0\,, \qquad \lf[\hat{p}_i,\hat{p}_j\ri]=\theta(\hat{x}^2)\hat{l}_{ji}\,, \qquad \lf[\hat{x}^i,\hat{p}_j\ri]=i\lf(g(\hat{x}^2)\delta^i_j+\bar{g}(\hat{x}^2)\frac{\hat{x}^i\hat{x}_j}{\hat{x}^2}\ri)\,,
\ee
with $g,$ $\bar{g}$ and $\theta$ 
%arbitrary [The functions cannot be completely arbitrary since they have to fulfill Eq.\eqref{rel2} for the left eqs in \eqref{eup} to be valid.]
being smooth functions of $\hat{x}^2=\hat{x}^i\hat{x}^j\delta_{ij}$ (where $\delta_{ij}$ denotes the Kronecker delta) and the deformed generator of rotations $\hat{l}_{ij}=2\hat{x}_{[i}\hat{p}_{j]}$. Note that recovery of the Heisenberg algebra for $\hat{x}\rightarrow 0$ requires $\lim_{\hat{x}\to0}g=1,$ $\lim_{\hat{x}\to0}\bar{g}=\lim_{\hat{x}\to0}\theta=0.$
Furthermore, since the operators $\hat{x}^i$ and $\hat{p}_j$ must satisfy Jacobi identities, the characteristic functions are not independent. Due to Eqs.~\eqref{eup}, the only non-trivial identity yields 
\be\label{jacobi}
    \lf[\hat{p}_i,\lf[\hat{p}_j,\hat{x}^k\ri]\ri] + \lf[\hat{p}_j,\lf[\hat{x}^k,\hat{p}_i\ri]\ri] + \lf[\hat{x}^k,\lf[\hat{p}_i,\hat{p}_j\ri]\ri]=0\,.
\ee
By virtue of the relation 
\be\label{rel}
    \lf[g(\hat{x}^2),\hat{p}_i\ri]=2i\lf(g(\hat{x}^2)+\bar{g}(\hat{x}^2)\ri)g'(\hat{x}^2)\hat{x}_i\,,
\ee
which holds also for $\bar{g}(\hat{x}^2)$ and where $'$ denotes a derivative with respect to $\hat{x}^2$, it is possible to show that 
Eq. \eqref{jacobi} reduces to
\be\label{mid}
%    2\lf(g(\hat{x}^2)+\bar{g}(\hat{x}^2)\ri)g'(\hat{x}^2)\hat{x}^i\delta^j_k+\frac{g(\hat{x}^2)\bar{g}(\hat{x}^2)}{\hat{x}^2}\hat{x}^j\delta^i_k=2\lf(g(\hat{x}^2)+\bar{g}(\hat{x}^2)\ri)g'(\hat{x}^2)\hat{x}^j\delta^i_k+\frac{g(\hat{x}^2)\bar{g}(\hat{x}^2)}{\hat{x}^2}\hat{x}^i\delta^j_k\,.
    \theta(\hat{x}^2) [\hat{x}^k , \hat{l}{_{ji}}]
    = \left[2 (g(\hat{x}^2) + \bar{g}(\hat{x}^2)) g'(\hat{x}^2) - \frac{g(\hat{x}^2) \bar{g}(\hat{x}^2)}{\hat{x}^2}\right] (x_i \delta^k_j - x_j \delta^k_i).
\ee
The ensuing uncertainty relations can be obtained by invoking the Robertson-Schr\"odinger prescription, according to which the standard deviations of positions and momenta $\Delta x^i$ and $\Delta p_i$, respectively, satisfy
\be\label{rs}
\Delta x^i\Delta p_j\geq\frac{1}{2}\lf|\lf\langle\lf[\hat{x}^i,\hat{p}_j\ri]\ri\rangle\ri|=\frac{1}{2}\lf|\lf\langle g(\hat{x}^2)\ri\rangle\delta^i_j+\lf\langle\bar{g}(\hat{x}^2)\frac{\hat{x}^i\hat{x}_j}{\hat{x}^2}\ri\rangle\ri|\,.
\ee
To see how a minimal momentum scale emerges from the previous expression, we have to explicitly select the form of the arbitrary functions $g$ and $\bar{g}$.
Since we expect the contributions quantifying the deviation from quantum mechanics to be small, we choose the most 
%well-known value 
common model~\cite{Hinrichsen:1995mf,Kempf:1996ss,Park:2007az,Mignemi:2009ji,Ong:2018nzk,Wagner:2022rjg}
\be\label{g}
g(\hat{x}^2)=1+\al\hat{x}^2,\qquad\bar{g}(\hat{x}^2)=2\alpha\hat{x}^2,
\ee 
with $\al$ having the dimension of an inverse squared length, and where we consider states satisfying $|\al|\langle\hat{x}^2\rangle\ll1$.
Typically, due to the fact that corrections in common models have a cosmological origin (\ie, $\al\sim \pm\Lambda\sim \pm10^{-52}$ m$^{-2}$, with $\Lambda$ being the cosmological constant \cite{Park:2007az,Ong:2018nzk,Mignemi:2009ji}), they can be treated perturbatively.
The negative (positive) sign for the deformation parameter is associated with a universe that behaves like a (anti-)de Sitter space \cite{Bambi:2007ty,Ghosh:2009hp}. 

If one momentarily suspends Einstein's convention for implicit summation over repeated indexes and considers the case $i=j$ in Eq.~\eqref{rs}, for the model \eqref{g} it is easy to show that
\be\label{ur}
\Delta x^i\Delta p_i\geq\frac{1}{2}\lf(1+3\al(\Delta x^i)^2\ri)\,.
\ee
Equation~\eqref{ur} either entails the presence of a globally minimal momentum $\Delta p_i^{(\mathrm{min})}=\sqrt{3\al}$ if $\alpha>0,$ (\ie, of the order of the Gibbons-Hawking temperature), or a globally maximal length $\Delta x^i_{(\mathrm{max})}=1/\sqrt{3|\alpha|}$ if $\alpha<0,$ (\ie, of the order of the distance to the Hubble horizon). 

\section{Deformed observables and quantum nonlocality}

In order to investigate quantum nonlocality for a bipartite quantum system, we first have to illustrate the difference between physical and auxiliary observables and how to transform the latter into the former when minimal momentum is accounted for. By means of this distinction, the generalization of the CHSH inequalities in the context of the EUP becomes straightforward. In passing, the generalization of the angular momentum operator is also explored.

\subsection{Physical and auxiliary observables}

The operators $\hat{x}^i$ and $\hat{p}_j$ introduced in Eq.~\eqref{eup} represent \emph{physical} observables satisfying the deformed canonical commutation relations (DCCRs); this is required for the minimal length/maximal momentum scales to apply to actual measurements.
%If we want to recover standard quantum mechanics, we should seek position and momentum operators $\hat{X}^i$ and $\hat{P}_j$ such that 
%
%\be\label{ccr}
%\lf[\hat{X}^i,\hat{X}^j\ri]=\lf[\hat{P}_i,\hat{P}_j\ri]=0\,, \qquad \lf[\hat{X}^i,\hat{P}_j\ri]=i\delta^i_j\,,
%\ee
%which is realized when $g(\hat{x}^2)=1$ and $\bar{g}(\hat{x}^2)=0$. Starting from the DCCRs, it is straightforward to verify that, by suitably defining
%\be\label{aux}
%\hat{X}^i\equiv\frac{\hat{x}^i}{g(\hat{x}^2)}\,, \qquad \hat{P}_j\equiv\hat{p}_j\,,
%\ee
%the the new phase space variables $\hat{X}^i$ and $\hat{P}_i$ obey the undeformed Heisenberg algebra \eqref{ccr}. These operators are \emph{auxiliary} observables, to which the physical ones reduce in the quantum mechanical limit. 
Nonetheless, it is convenient to introduce a new set of operators which will be referred to as ``auxiliary variables'' and that satisfy the ordinary Heisenberg algebra
\be\label{ccr}
\lf[\hat{X}^i,\hat{X}^j\ri]=\lf[\hat{P}_i,\hat{P}_j\ri]=0\,, \qquad \lf[\hat{X}^i,\hat{P}_j\ri]=i\delta^i_j\,.
\ee
There is an infinite number of ways of connecting these quantities to the physical variables introduced in Eq. \eqref{ccr} by a non-canonical transformation \cite{Bosso:2022vlz}. However, concerning the present paper, there is a preferred path to follow. As a matter of fact, our interest is prominently focused on the definition of a deformed spin operator with which to analyze deviations from the standard quantum mechanical picture when a Bell-like test involving spinors is studied. Since we will start from a relativistic description (\ie, Dirac equation), the position operator has to be carefully handled, as its definition is physically consistent only in the non-relativistic regime.

To begin with, it is instructive to study the modifications at the level of the angular momentum algebra. As done in \cite{Bosso:2016frs} for GUP-type deformations,
we can define the physical angular momentum operator as $\hat{l}_i=\varepsilon_{ijk}\hat{x}^j\hat{p}^k$.
Consequently, using the commutation relations~\eqref{eup}, one obtains
\be\label{ang}
\lf[\hat{l}_i,\hat{l}_j\ri]=i\varepsilon_{ijk}\hat{l}^kg(\hat{x}^2)\,.
\ee
Note that the ordering of the factors $\hat{l}^k$ and $g(\hat{x}^2)$ is irrelevant, because these two operators commute. Furthermore, one may also introduce the auxiliary angular momentum operator which fulfills the standard SO(3) algebra, that is
\be\label{aux2}
\hat{L}^i=\frac{\hat{l}^i}{g(\hat{x}^2)}\,.
\ee
Intuitively, a similar deformation is expected to occur also for the spin operator, even though it is not related to the dynamical degrees of freedom, but rather to internal ones.
Such a generalization has already been speculated about in~\cite{Bosso:2016frs}, but a 
physical argument supporting this occurrence has only recently appeared in literature in the context of the GUP~\cite{Bosso:2022ogb}. Here, we will follow the same reasoning after the application of a EUP-type deformation.

First of all, since the aforementioned reasoning requires a non-relativistic limit of the Dirac equation, it is important to bear in mind that positions and momenta are not treated on the same footing in relativistic quantum mechanics. Therefore, it is unclear what a transformation of the position operator would imply at the level of the Dirac equation. Hence, it is preferable to encode the EUP-induced deformation into the transformation from the physical to the auxiliary momentum operator only, while physical and auxiliary positions are equal. This can be achieved via the transformation
\begin{equation}
    \hat{x}^i\equiv \hat{X}^i\qquad \hat{p}_i\equiv g(\hat{X}^2) \hat{P}_i+\bar{g}(\hat{X}^2)\frac{\hat{X}^i\hat{X}^j}{\hat{X}^2}\hat{P}_j,\label{aux}
\end{equation}
as it can be verified using Eqs. \eqref{eup}.
It is worth observing that, in the Heisenberg-limit ($g=1,$ $\bar{g}=0$), the two sets of operators (\ie, physical and auxiliary) coincide.

Expressed in terms of the auxiliary momenta (whose position representation are the familiar partial derivatives), we obtain the deformed Dirac equation describing a fermion of charge $e$ nonminimally coupled to a background vector potential $A_i$ as
\begin{equation}
    i\partial_t\psi=\left[\gamma^i\left(p_i(P,X)-eA_i\right)+\gamma^0m\right]\psi \,, 
    \label{eqn:Dirac_physical}
\end{equation}
with the Dirac matrices $\gamma^\mu$ (Dirac representation is understood). Next, splitting up the Dirac spinor into particle $\varphi$ and antiparticle $\chi$ as $\psi=(\varphi,\chi),$ the non-relativistic limit of this equation can be obtained along the line of \cite{Bosso:2022ogb}. While the antifermion becomes a non-propagating degree of freedom, the fermion satisfies the effective Schr\"odinger equation
\begin{equation}
    i\partial_t\varphi=\frac{1}{2m}\left[\sigma^i\left(p_i(P,X)-eA_i\right)\right]^2\varphi \, ,
    \label{eqn:Schrod_physical}
\end{equation}
with the Pauli matrices $\sigma^i.$ The interesting quantity to consider for the purpose of the present study is the part of the underlying Hamiltonian which couples to the magnetic field $B^i\equiv\varepsilon^{ijk}\nabla_jA_k,$ as this is how we identify the physical spin operator. In particular, the achieved coupling turns out to be
\begin{equation}
    (\hat{l}_i+g\sigma_i)B^i\equiv(\hat{l}_i+2\hat{s}_i)B^i,
\end{equation}
which is consistent with the result obtained in \cite{Bosso:2022ogb} for the GUP and with Eq. \eqref{aux2}.

Indeed, by introducing the auxiliary spin operator furnishing a three-dimensional representation of $SU(2)$ as $\hat{S}^i=\sigma^i/2,$ one obtains the relation
\be\label{aux3}
\hat{S}^i=\frac{\hat{s}^i}{g(\hat{x}^2)}\,.
\ee
%where in Ref.~\cite{Bosso:2022ogb} the function $g$ depends on $\hat{p}^2.$ Nevertheless, the argument can be straightforwardly reiterated so as to include the EUP-involving deformation to deduce the above expression. 
%Equation~\eqref{aux3} 
Thus, physical and auxiliary angular momenta as well as spins satisfy the same algebras as it could have been intuitively expected \emph{a priori}.

\subsection{CHSH experiment}

In order to quantify the deformations to measurements of quantum nonlocality, we rely on the CHSH version~\cite{Clauser:1969ny} of Bell theorem~\cite{Bell:1964kc}. For this purpose, one needs to identify two observers, Alice (A) and Bob (B), performing independent measurements on distinct parts of a bipartite quantum system. The separation between A and B must be space-like to avoid any causal connection between the two outcomes. Suppose $\hat{A}$ and $\hat{B}$ are the dichotomous observables associated to Alice and Bob, respectively, whilst $(a,a')$ and $(b,b')$ denote two different settings available for the detectors probing $\hat{A}$ and $\hat{B}$. Because of the absence of causal interactions between the parties, there is no dependence of $\hat{A}$ on $(b,b')$ and vice-versa, which means that the only possible configurations are $\hat{A}(a)$, $\hat{A}(a')$, $\hat{B}(b)$ and $\hat{B}(b')$. By relying on a classical interpretation of probability while suitably combining all correlation functions $C((a,a'),(b,b'))$ achievable by changing the detectors' settings, it is possible to show that~\cite{Clauser:1969ny}
\begin{equation}\label{s}
    \mathcal{S}=\big|C\left(a,b\right)-C\left(a,b'\right)+C\left(a',b\right)+C\left(a',b'\right)\big|\leq2 .  
\end{equation}
In the context of quantum mechanics, starting from the state $\rho_{AB}$ on which the measurements are performed, the correlation functions are given by 
\be\label{cf}
C(a,b)=\mathrm{Tr}\lf[\rho_{AB}\lf(\hat{A}(a)\otimes\hat{B}(b)\ri)\ri]\,,
\ee 
and similarly for the other combinations of $(a,a')$ and $(b,b')$.

Let us now assume that an experiment involving two spin-1/2 particles is performed; since there are only two available degrees of freedom per particle, one can choose the following observables:
\begin{align}\label{obs}
    \hat{A}(a) &= 2\hat{s}_3\otimes\id\,, & 
    \hat{A}(a') &= 2\hat{s}_1\otimes\id\,, &
    \hat{B}(b) &= -\frac{\id\otimes2\left(\hat{s}_3+\hat{s}_1\right)}{\sqrt{2}}\,, & 
    \hat{B}(b') &= \frac{\id\otimes2\left(\hat{s}_3-\hat{s}_1\right)}{\sqrt{2}}\,,
\end{align}
where $\id$ is the $2\times2$ identity.

Consider the undeformed case first, for which it is sufficient to consider a generic two-qubit state.  The corresponding density matrix can be written as~\cite{chuang} 
\be\label{gen}
\rho_{AB}=\frac{1}{4}\sum_{i,j=0}^3\al_{ij}{\si}_i\otimes\si_j\,, \qquad \sigma_0\equiv\id.
\ee
As a result, the sum of correlation functions \eqref{s} yields $\mathcal{S}=\sqrt{2}\,|\al_{11}+\al_{33}|$, which still satisfies the inequality~\eqref{s} as long as \mbox{$|\al_{11}+\al_{33}|\leq\sqrt{2}$}.
However, to provide a meaningful example, one can consider Bell diagonal states, whose general expression reads
\be\label{bds}
\rho_{AB}=p_0|\Phi^+\rangle\langle\Phi^+|+p_1|\Psi^+\rangle\langle\Psi^+|+p_2|\Psi^-\rangle\langle\Psi^-|+p_3|\Phi^-\rangle\langle\Phi^-|\,, \qquad \sum_{i=0}^3p_i=1\,,
\ee
and where $|\Phi^\pm\rangle$, $|\Psi^\pm\rangle$ represent the maximally entangled Bell states
\be\label{bs}
|\Phi^\pm\rangle=\frac{1}{\sqrt{2}}\lf(|00\rangle\pm|11\rangle\ri)\,, \qquad |\Psi^\pm\rangle=\frac{1}{\sqrt{2}}\lf(|01\rangle\pm|10\rangle\ri)\,.
\ee
Here, $|0\rangle$ and $|1\rangle$ stand for the eigenvectors of $\hat{S}_3$ with eigenvalues $\pm1$, respectively. By means of the density matrix $\rho_{AB}$ appearing in~\eqref{bds}, we obtain $\mathcal{S}=2\sqrt{2}\,|p_2-p_0|$. For either $p_0=1$ or $p_2=1$ (\ie, when a pure Bell state is considered), one recovers
\be\label{tsirelson}
\mathcal{S}_{\mathrm{max}}=2\sqrt{2}\,,
\ee 
which is known as Tsirelson bound~\cite{tsir}. This is the highest value that can be achieved by violating the CHSH inequality in the framework of quantum mechanics, thereby setting a fundamental limit to the allowed degree of quantum nonlocality. It is believed that this constraint may be overcome when quantum gravitational effects are accounted for~\cite{Bosso:2022ogb}, but in what follows it will be proved that also the limitation related to the existence of a minimal momentum or a maximal length affects $\mathcal{S}_{\mathrm{max}}$. Henceforth, we focus on the case in which $p_2=1$, thus dealing with an overall spin singlet state that can be identified with the Bell state $|\Psi^-\rangle=(|01\rangle-|10\rangle)/\sqrt{2}$.

Now, assuming to work with non-trivial deformations, the physical operators act not only on the qubit-Hilbert space, but also on the particles' positional states. This occurs because the initial state $|\psi_{AB}\rangle$ utilized to define the pure-state density matrix $\rho_{AB}=|\psi_{AB}\rangle\langle\psi_{AB}|$ can always be split as 
\be\label{fact}
|\psi_{AB}\rangle=|\psi_x\rangle\otimes|\psi_s\rangle\,,
\ee
with $|\psi_x\rangle$ carrying the dependence on the position and $|\psi_s\rangle$ encoding the information associated with the spin. In the standard quantum mechanical scenario, the first term is factored out, as the observables only act on the space of the spin degree of freedom, thereby leaving $\langle\psi_x|\psi_x\rangle=1$.
On the other hand, once the EUP is imposed, each operator in Eqs.~\eqref{obs}
can be written in terms of the operators $\hat{S}_i$ following the ordinary $SO(3)$ algebra and
a further term $g(\hat{x}^2)$ contributing to its expectation value in the form
\be\label{gexp}
\langle\psi_x|g(\hat{x}^2)|\psi_x\rangle=\langle g\rangle\,,
\ee
for both $\hat{A}$ and $\hat{B}$, whilst the spin-related part is left untouched. This occurs because, in the EUP framework, the physical spin is no longer $\hat{S}_i$, but rather $\hat{s}_i$ as defined in Eq. \eqref{aux3}.

Putting all these results together, it can be shown that the Tsirelson bound~\eqref{tsirelson} is modified as 
\be\label{fin}
\mathcal{S}^{\mathrm{EUP}}_{\mathrm{max}}=2\sqrt{2}\, \langle{g(\hat{x}_A^2)}\rangle \langle{g(\hat{x}_B^2)}\rangle\,,
\ee
where we used that the positional Hilbert spaces of the two particles factorize (their respective position operators are differentiated by the subscripts $A$ and $B$). By using a perturbative expansion of Eq.~\eqref{g} to render the dependence on the deformation parameter $\al$ explicit, one finds that
\be\label{fin2}
\mathcal{S}^{\mathrm{EUP}}_{\mathrm{max}}=2\sqrt{2}\big[1+\al\lf(\langle \hat{x}_A^2\rangle+\langle \hat{x}_B^2\rangle\ri)\big]\,.
\ee
To provide a concrete example, the following assumptions will be considered: in light of the reasoning outlined in Sec.~II when trying to mimic the cosmological horizon, $\al$ is chosen to be of the order $\mathcal{O}(\Lambda)$. As the correction to the standard Tsirelson bound in~\eqref{fin2} is expected to be extremely tiny, large values of $\langle g(\hat{x}^2_A)\rangle$ and  $\langle g(\hat{x}^2_B)\rangle$ are needed to make it amenable to detection. Additionally, as mentioned in the introduction, it has been argued \cite{Bambi:2007ty,Ghosh:2009hp} that the cosmological EUP (with positive cosmological constant) entails a negative parameter $\al,$ \ie a maximal length.

Furthermore, let us consider the problem to be viewed from the quantum reference frame of particle $A.$ Under this assumption, the positional state of $A$ is sharply peaked in the origin such that $\langle \hat{x}^2_A\rangle\simeq 0.$ Then, it is easy to show that
\be\label{fin3}
\mathcal{S}^{\mathrm{EUP}}_{\mathrm{max}}\simeq2\sqrt{2}\big(1-|\al|  \langle x_{B}^2\rangle\big)\,.
\ee
As a result, the deformation of the Tsirelson bound is always smaller than the quantum mechanical constraint $\mathcal{S}_{\mathrm{max}}$.

Thus, a maximal length implies the degradation of quantum nonlocal correlations shared by two parties of a bipartite system. As particle $B$ recedes from the origin, the amount of nonlocality keeps decreasing, until it eventually becomes smaller than $2$. After this threshold, the CHSH inequality~\eqref{s} holds true again, signaling a transition from a quantum to a ``classical'' behavior of the correlations. However, it must be stressed that the deformation also affects classical correlations, thus leading to an analogous degradation. 
This phenomenon is expected to be relevant for distances between the two spins greater than or equal to
\be\label{xab}
\sqrt{\langle x^{2}_{B}\rangle}\sim 10^{25} \mathrm{m}\,,
\ee
which is comparable with the magnitude of the radius of the cosmological horizon.

\section{Discussion}

As recently pointed out in the literature, modifications of the Heisenberg algebra imply a redefinition of the concept of physical spin \cite{Bosso:2022ogb}. The present paper corroborates this finding at the level of general position-dependent corrections, which entail a EUP-deformed quantum mechanics featuring either a maximal length or a minimal momentum. In particular, such a result can be achieved starting from the Dirac equation in the presence of electromagnetic interaction and taking its non-relativistic limit. In so doing, the resulting modified Schrödinger equation includes a coupling of internal degrees of freedom to the magnetic field that clearly denotes  a deformed version of the $SU(2)$-algebra fulfilled by the physical spin operator. This modification is in accordance with changes to the orbital angular momentum algebra already observed in earlier works \cite{Benczik:2002tt,Bosso:2016frs,Lake:2019nmn}. 

As a result of the aforesaid deformation, motional degrees of freedom of a quantum state cannot be factored out when measuring physical spins. Correspondingly, the degree of nonlocal correlations shared by bipartite quantum systems is non-trivially affected. For instance, in models containing a fundamental maximum length, the Tsirelson bound $\mathcal{S}_{\mathrm{max}}$ (which limits
the highest achievable amount of quantum mechanical nonlocality) does not apply in general. Indeed, it turns out that its generalization $\mathcal{S}^{\mathrm{EUP}}_{\mathrm{max}}$ is a monotonically decreasing function of the distance between the two parties which coincides with $\mathcal{S}_\mathrm{max}$ in the limit of vanishing separation. Assuming the maximal length to be of the order of the radius of the cosmological horizon, this allows for the emergence of an effectively classical behavior at cosmological scales.

Of course, nonlocal correlations are but one example of the wide applicability of the deformation of spin operators found here and in Ref. \cite{Bosso:2022ogb}. More detailed analyses of its consequences will be addressed in future works.

\section*{Acknowledgement}

The authors acknowledge support by MUR (Ministero dell'Universit\`a e della Ricerca) via the project PRIN 2017 ``Taming complexity via QUantum Strategies: a Hybrid Integrated Photonic approach'' (QUSHIP) Id. 2017SRNBRK.  P.B., F.W. and L.P. acknowledge networking support by the COST Action CA18108. L.P. is grateful to the ``Angelo Della Riccia'' foundation for the awarded fellowship received to support the study at Universit\"at Ulm.

%\bibliographystyle{utphys}
%\bibliography{ref.bib}

\providecommand{\href}[2]{#2}\begingroup\raggedright\endgroup
\end{document}